Letter

# Short-term Effects of Gamma Ray Bursts on Earth

Osmel Martín[1], Douglas Galante[2], Rolando Cárdenas[3] and J.E. Horvath[4]

[1, 3] Department of Physics, Universidad Central de Las Villas, Santa Clara, Cuba.
Phone 53 42 281109 Fax 53 42 281109 e-addresses: [1] osmel@uclv.edu.cu; [3] rcardenas@uclv.edu.cu

[2, 4] Department of Astronomy, Instituto de Astronomia, Geofísica e Ciências Atmosféricas, Universidade de São Paulo, Brazil. e-addresses: [2] douglas@astro.iag.usp.br; [4] foton@astro.iag.usp.br

**Abstract:** The aim of the present work is to study the potential short-term atmospheric and biospheric influence of Gamma Ray Bursts on the Earth. We focus in the ultraviolet flash at planet's surface, which occurs as a result of the retransmission of the γ radiation through the atmosphere. This would be the only important short-term effect on life. We mostly consider Archean and Proterozoic eons, and for completeness we also comment on the Phanerozoic. Therefore, in our study we consider atmospheres with oxygen levels ranging from $10^{-5}$ to 1 of the present atmospheric level, representing different moments in the oxygen rise history. Ecological consequences and some strategies to estimate their importance are outlined.

## I - Introduction

During the Archean and Proterozoic eons the atmosphere of our planet evolved from very low oxygen content, perhaps lower than $10^{-5}$ of the present atmospheric level (pal) of $O_2$, towards our modern Phanerozoic atmosphere with roughly 20 % $O_2$. Many authors consider that the $O_2$ rise in the Earth's atmosphere is intrinsically linked with the development of the first photosynthetic bacteria and with the termination of the chemical sinks which consumed it (Catling & Clair, 2005; Catling, Zahnle & McKay, 2001). Although many details on the evolution of our atmosphere remain controversial (Cockell, 2000; Cockell & Raven, 2007), several authors accept that the irreversible increase of $O_2$ took place around 2.3 billion of years ago, millions of years after the first evidences of photosynthetic life appeared in a global extension. Seemingly, this process was characterized by an abrupt increase of the $O_2$ level from around $10^{-5}$ $O_2$ pal to around $10^{-1}$ $O_2$ pal in about 100 million of years. After this initial sudden rise, the oxygen levels have increased (with several fluctuations) up to the present value. Accordingly, we considered in the present study atmospheres with oxygen levels ranging from $10^{-5}$ to 1 of the present atmospheric level (pal), representing different moments in the oxygen rise history.

As we said above, many issues of the atmospheric composition and evolution during the Archean and Proterozoic are nowadays controversial. It is a big interdisciplinary challenge to make exact quantitative estimations on several issues in these eons. However, roughly speaking, we can consider that our studied atmospheres correspond with a reasonable approximation to past atmospheres and geologic times as we state in the Table 1 below.

| Oxygen content in present atmospheric level ($O_2$ pal) | Geologic time |
|---|---|
| 1 | Late Proterozoic era and Phanerozoic eon (0.8 Ga ago up to the present) |
| $10^{-1}$ or $10^{-2}$ | Mid-Proterozoic era (2.3 Ga – 0.8 Ga before present) |
| $10^{-3}$, $10^{-4}$ or $10^{-5}$ | Archean eon and Early Proterozoic era (3.8 Ga – 2.5 Ga) |

**Table 1 Model atmospheres studied and the geologic time in which they probably existed**

During long periods of time, stochastic variation of the UV levels reaching the surface of Earth could have significantly influenced the evolution of life. Several astrophysical sources of radiation have been proposed, such as solar flares, supernovae explosions and Gamma-Ray Bursts (GRB) (Thorsett, 1995; Smith, Scalo & Wheeler, 2004a; Thomas et al., 2005; Gehrels et al, 2003). For Earth-like planets with an atmospheric density $\rho \gg 100$ g/cm$^3$, the harmful influence is not directly associated to the primary ionizing flux, because the latter is strongly absorbed in the upper atmosphere. The most immediate effect on surface is the large UV reemission flux resulting from the interaction of primary radiation with the atmospheric constituents. This effect is usually called "UV-flash" and its biological relevance is related today mainly to the presence of UV blocking agents in the atmosphere, especially in the region of 200-310 nm. In this regard, oxygen plays a dual role in the atmosphere; the first is associated directly with its intrinsic absorption features mainly in the UV-C (100-280nm) band and the second in the ozone layer formation, which strongly absorbs in the UV-B (280-315nm) region.

However, the irradiation by energetic astrophysical sources could cause a massive depletion of the ozone layer resulting in an increase on the solar UV flux reaching the ground. This is likely to be the major impact on the UV radiation flux at a planet's surface if a significant ozone layer is present, as in today's Earth. The high-energy gamma rays from the burst should break apart $N_2$ and $O_2$, creating nitrogen oxides ($NO_x$), which then catalyze ozone destruction. This, in turn, would allow more of the solar UV radiation to reach the planet's surface, which could remain enhanced for several years, until ozone layer recovers. Another planetary effects are likely to occur, and in general for a detailed modeling of the potential action of a GRB on the Phanerozoic Earth we recommend (Thomas et al, 2005) and references therein, while for a compact synthesis the work of Galante & Horvath (2007) may be useful.

On the other hand, the atmosphere in Archean and Early-Proterozoic Earth presented low oxygen and, consequently, low ozone levels. A strong ozone UV shield, such as the one our planet had in the Phanerozoic eon, was absent. Thus, in our modeling, we have chosen not to take into account the effect of ozone depletion, and focused on the effect of UV retransmission only, the above mentioned "UV-flash". Basically, this would be the only important short-term effect on life. Warming of the atmosphere and alterations of its chemistry, with potential climatological and biological consequences, would be rather long-term influences. The detailed modeling of long-term effects of a GRB on these geologic times would entail a huge interdisciplinary challenge (Catling, 2008), and it is well beyond the scope of this study.

On the other hand, it has been suggested that the probability of direct effects from astrophysical events such as supernovae explosions or GRB's were potentially bigger during the early Earth (Scalo & Wheeler, 2002), namely when atmospheric composition, climate and biosphere structure differed considerable from the present one. Events like GRB's should have had different consequences during the (seemingly) less protected life in the Archean or Early-Proterozoic Earth. No other efficient candidate to replace the ozone/oxygen screening properties in the atmosphere has been firmly established (Pavlov, Brown & Kasting, 2001).

It can be argued that he retransmitted UV would play even a more important role in planets orbiting M stars, where flares are common in the first billion years of the star's life. These stars are the most common in our Galaxy and, if they are parent stars of planetary systems bearing life, they could induce stochastically varying mutation rates and frequent hypermutation episodes (Smith, Scalo & Wheeler, 2004b).

**II - Basic assumptions**

A – <u>Spectrum of the Gamma Ray Burst</u>

The assumed broken power law for the GRB spectrum, and the parameterizations used in this work, were taken directly from (Smith, Scalo & Wheeler, 2004a):

$$\frac{dN}{dE} = k\left(\frac{E}{100 keV}\right)^{\alpha} \exp\left(-E/E_0\right) \quad \text{for} \quad E \leq (\alpha - \beta)E_0, \qquad (1)$$

$$\frac{dN}{dE} = k\left[\frac{(\alpha - \beta)E_0}{100 keV}\right]^{\alpha - \beta} e^{\beta - \alpha}\left(\frac{E}{100 keV}\right)^{\beta} \quad \text{for} \quad E \geq (\alpha - \beta)E_0, \qquad (2)$$

where $E_0 = 250$ keV, $\alpha = -0.9$ and $\beta = -2.3$

The energy range considered was 50 keV < E < 3 MeV. The original spectrum was split in 100 equal logarithmically spaced "monochromatic" energy bins with a corresponding value of mean energy and flux associated to each bin.

B - <u>Atmospheric Models</u>

For the atmospheric models, an exponential and well-mixed (except the ozone) atmosphere was considered. The main parameters used were

- Scale Height (Ho) = 8 km
- Density ($\rho$) = $1.3 \times 10^{-3}$ g/cm$^3$

First, we calculate the γ energy deposition in the atmosphere and then, using these results, we compute the UV reemission. The actual composition of the atmosphere is quite irrelevant for γ energy deposition studies, thus it was fixed to the present value without substantial error. However, atmospheric constituents are important for the UV reemission process and its consequent biological implications. Therefore, to estimate the UV radiation levels reaching the ground, our six different atmospheric compositions were used, respectively with $10^{-5}$, $10^{-4}$, $10^{-3}$, $10^{-2}$, $10^{-1}$ and $10^{0}$ $O_2$ pal. We then follow (Segura et al., 2003) in other aspects of our atmospheric models:

- We adopted their data for the ozone column.
- Ground atmospheric pressure was set to 1 atm.
- We allowed $CO_2$ to vary to fill in the rest of the atmosphere as oxygen is removed. This is important because $CO_2$ was a dominant candidate gas in ancient environments and its spectroscopic characteristics can attenuate the radiation levels mainly in UV-C region (Cockell, 2000; Cnossen et al, 2007).

To implement the ozone column in each case, two procedures were applied. The first was a simple direct rescaling of the *standard USA 76 atmosphere*, and the second was a shift down of the entire ozone profile and then a rescaling of it to the appropriate case. It is well known that depletion of $O_2$ tends not only to make the ozone layer thinner, but also closer to the troposphere (Segura et al, 2003; Kasting &Catling, 2003). The second procedure attempts in a very rough way to mimic the latter effect.

C - Interactions of Gamma rays with the atmosphere

To study the interactions of γ rays in the atmosphere we employed the general procedure described in Gehrels et al (2003). The atmosphere was splitted in 90 layers from 0 to 180 km and the wavelength range was splitted in 100 beams. Each normal incident energy beam was attenuated by an exponential Beer's law of the form $N_{i,j} = N_i^0 e^{-\mu_i x_j}$ according to its own flux and mean energy, where $N_{i,j}$ is the photon flux remaining from the incident monoenergetic beam $N_i^0$; $x_j$ is the column density (in g/cm$^2$) of the layer and $\mu_i$ is the corresponding mass attenuation coefficients taken from (Berger, M. et al, 2005). The energy deposited at each layer was determined using the difference between the total fluxes incident on adjacent layers:

$$Fdep,i = \sum_{j=1}^{100} \left(N_{i,j} - N_{i,j-1}\right) \qquad (3)$$

For the UV-reemission treatment it was assumed that the emission follows roughly the law (Smith, Scalo & Wheeler, 2004a):

$$\frac{dF_{UV,i}}{d\lambda} = \frac{Fdep,i}{\lambda \ln\left(\lambda_{max}/\lambda_{min}\right)} , \qquad (4)$$

where $Fdep,i$ is the total energy deposited in the *i-th* layer. The values $\lambda_{max}$ and $\lambda_{min}$ are associated to the typical limits of nitrogen bands and they were set to 130 and 600 nm respectively.

To compute the incident UV radiation fluxes on the surface of Earth, both photoabsorption and Rayleigh scattering were taken into account. The net effect at surface was computed as the sum of the emission effects of each layer.

D – Biological Effects and critical Distances

In this section we show how to calculate the critical distances of the GRB progenitors which would cause significant damages on the biota. To do this, we followed (Cockell, 2000; Cockell & Raven, 2007), i. e., we use the action spectrum $e(\lambda)$ for DNA damage, convolve it with the incident UV flux density $E(\lambda)$, and calculate the effective biological irradiances or fluxes $E^*$:

$$E^* = \sum e(\lambda)E(\lambda)\Delta\lambda \qquad (5)$$

We then define and calculate the effective biological fluences $F^*$:

$$F^* = E^*\Delta t, \qquad (6)$$

where $\Delta t$ is the exposure time to the UV radiation.

In what follows, the subscript *GRB* will refer to the retransmitted UV radiation coming from the Gamma Ray Burst, and the subscript *Sun* will stand for the UV coming from the Sun. There is much variability between different species concerning their resistance to radiations, and therefore to define *significant biological damage* is a fuzzy problem. In our case, we modified the criterion used in (Thomas et al, 2005) for the action of a GRB on the Phanerozoic eon. Instead of using biological effective irradiances we introduce *effective biological fluences* (eq. 6), to describe the potential biological damage of UVR. We then assume the following condition for having significant marine mortality under the action of the UV flash:

$$F^*_{GRB} = nF^*_{Sun} \qquad (7)$$

This means that in the exposure time of ten seconds to the UV flash, the organism would be exposed to an effective biological fluence which is *n* times greater than the one they are used to receive in the *whole day* from the Sun. It seems reasonable to expect significant mortality of surface inhabitants of both the sea and land already for *n =1*, especially since in ten seconds the repair mechanisms will have very little time to act against such a huge effective biological fluence. Of course, mortality should not be expected to occur *only* during the exposure time, but also later as irreversible sequels take over. Another issue is the duration of a day in different eons. Although uncertainties remain concerning this, it seems that the length of the day in our planet has increased from some 15 hours in Early Archean up to our present 24 hours. We have adopted a median value of 20 hours and finally use its half (10 hours) for the actual daily exposition to the Sun. Differences in exposure time due to latitude or specific eon do not change appreciably the main results of this work.

To proceed we define a dimensionless effective biological irradiance $E^*_{ad}$

$$E^*_{ad} = \frac{E^*_{GRB}}{I^{top}_{GRB}} \quad , \tag{8}$$

where $I^{top}_{GRB}$ is the full flux of the GRB at the top of the atmosphere. Combining eqs. (6-8) we obtain

$$E^*_{ad} I^{top}_{GRB} \Delta t_{GRB} = n E^*_{Sun} \Delta t_{Sun} \tag{9}$$

The fluence of the GRB at the top of the atmosphere is

$$I^{top}_{GRB} \Delta t_{GRB} = \frac{E_\gamma}{\Delta\Omega \times D^2} \tag{10}$$

In the above formula $E_\gamma = 5 \times 10^{43}$ J is the gamma energy, $\Delta\Omega$ is the solid angle fixed to 0.01 sr according to (Frail et al., 2001), and $D$ is the distance of the progenitor of the GRB. The absorption by the interstellar medium was not included in our analysis because it is almost completely transparent to high-energy photons.

Substituting eq. (10) into eq. (9) and solving for $D$ we obtain an expression to calculate the critical distances from which a GRB emission would cause significant surface mortality, as the condition given in eq. (7) is fulfilled

$$D = \sqrt{\frac{E_\gamma \times E^*_{ad}}{\Delta\Omega \times \Delta t_{Sun} \times n \times E^*_{Sun}}} \tag{11}$$

An issue here is to estimate $E^*_{Sun}$ for our different atmospheres. We propose the following approximate procedure: we consider that the Archean atmosphere studied in (Cockell, 2000) corresponds to our $10^{-5}$ pal $O_2$ atmosphere. From Fig. 5 of that paper it can be inferred approximate values of 80 W/m$^2$ and 20 W/m$^2$ for solar zenithal angles of 0° and 60° respectively. As a rough average we used $E^*_{Sun} = 50$ W/m$^2$. Then we use Table 3 of (Segura et al, 2003), where $E^*_{Sun}$ are normalized to the present value (in that reference paper the $E^*_{Sun}$ have dimensions and are shown in Table 2). In that reference the authors only used the wavelength range (295-315) nm; so we are using this range as a proxy for all UV effects. This will imply some underestimation of $E^*_{Sun}$ because of the ozone absorption of the UV-C band in the range (200-280) nm. Therefore, the lesser the oxygen content of the atmosphere, the higher will be that underestimation. However, we consider that the errors introduced will be relatively small.

**III – Results**

A – Retransmitted ultraviolet radiation reaching the ground

The table below shows the UV energy reaching the ground, expressed as a fraction of the original incident gamma energy at the top of the atmosphere. In general, the fraction of the incident fluence which is retransmitted to UV is almost independent of the mean energy of the incident photons, because the number of secondary electrons available for molecule excitation only depends on the total energy of primary electrons. This last quantity depends on the total fluence of ionizing radiation, and not on the specific form of the spectrum of the GRB. Therefore, our reported values below are quite similar to those presented in Smith, Scalo and Wheler (2004b), actually a bit lower, as we replaced oxygen by $CO_2$ and not by $N_2$, as done in the mentioned reference.

| $O_2$ pal | UV-A 315-400 nm | UV-B 280-315 nm | UV-C 130-280 nm | TOTAL 130-600 nm |
|---|---|---|---|---|
| $10^0$    | $2.330 \times 10^{-02}$ | $6.311 \times 10^{-04}$ | $3.043 \times 10^{-12}$ | $2.022 \times 10^{-01}$ |
| $10^{-1}$ | $5.867 \times 10^{-02}$ | $2.789 \times 10^{-03}$ | $2.418 \times 10^{-09}$ | $2.480 \times 10^{-01}$ |
| $10^{-2}$ | $6.484 \times 10^{-02}$ | $5.161 \times 10^{-03}$ | $4.817 \times 10^{-07}$ | $2.583 \times 10^{-01}$ |
| $10^{-3}$ | $6.598 \times 10^{-02}$ | $8.943 \times 10^{-03}$ | $5.802 \times 10^{-05}$ | $2.643 \times 10^{-01}$ |
| $10^{-4}$ | $6.616 \times 10^{-02}$ | $1.187 \times 10^{-02}$ | $2.338 \times 10^{-03}$ | $2.699 \times 10^{-01}$ |
| $10^{-5}$ | $6.617 \times 10^{-02}$ | $1.210 \times 10^{-02}$ | $2.933 \times 10^{-03}$ | $2.707 \times 10^{-01}$ |

**Table 2: UV energy reaching the ground expressed as fraction of the original incident gamma energy at the top of the atmosphere.**

In order to make a better estimation of the biological damage of the GRB, we have also obtained the (more resolved) incident spectrum at the surface of the planet, as can be shown in Fig. 1. These UV irradiances are normalized to the full flux supplied by the GRB at the top of the atmosphere. The wavelengths below 200 nm were not considered, as they are almost totally absorbed in the atmosphere.

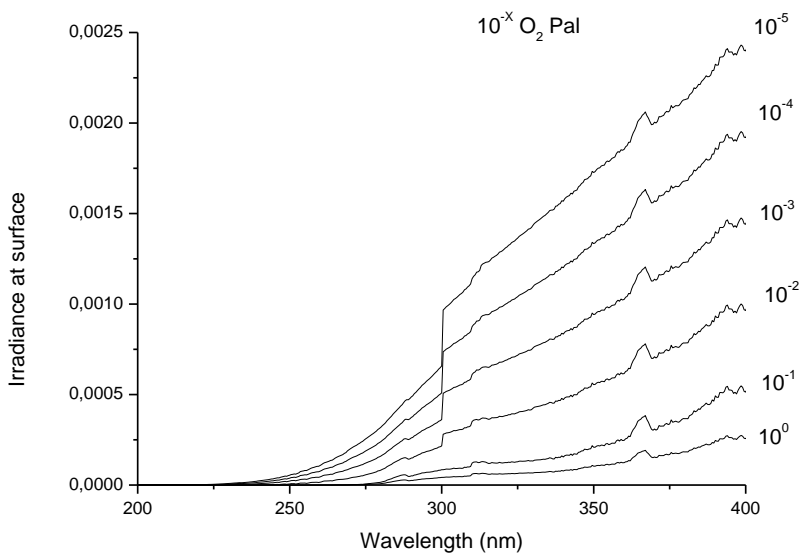

**Fig. 1 UV irradiances reaching the ground for different atmospheres and wavelengths, normalized to the full flux of the GRB at the top of the atmosphere.**

B – <u>Biological effective irradiances</u>

Applying the methodology described in subsection II-D, we convolved the above spectra with the action spectrum for DNA, to obtain the biological effective irradiances. Fig. 2 below shows the dimensionless irradiances for the studied atmospheres. As expected, the smaller the oxygen content, the greater the DNA damage caused by the GRB, as the atmosphere would have less ozone to shield the retransmitted UV.

We would like to comment on the relative importance of the flash here. Usually, the authors engaged in the modelling of the effects of GRB's on Earth tend to ignore the UV flash, largely based on the fact that the longer term effects are more important. In addition, in atmospheres having an oxygen-ozone shield, the UV-C band is almost totally absorbed, and its effective biological effect is relatively small, as can be seen in the two lower curves in Fig.2. However, with the decrease of the oxygen content, an important fraction of the very biologically damaging wavelengths below 280 nm pass through the atmosphere and reach the ground. Therefore, for atmospheres having from $10^{-3}$ to $10^{-5}$ pal (probably Archean and Early Proterozoic atmospheres), the peak of the effective biological irradiance is in the UV-C region, as can be seen in the upper three curves of the above figure. In our opinion, this makes worth to investigate the potential biological effects of the UV flash, especially for low-ozone atmospheres.

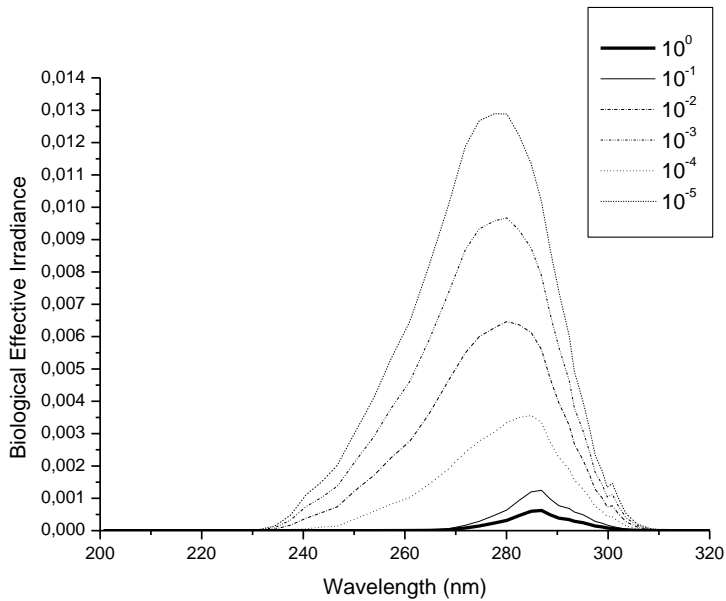

**Fig. 2 Biological effective irradiances normalised to the full flux of the GRB at the top of the studied atmospheres. As expected, the lesser the oxygen content, the greater the DNA damage caused by the GRB.**

C-Critical Distances

Applying the methodology described in subsection II-B, we have obtained the critical distances for our criteria for severe damage to the biota. We have several values for the parameter *n* and have denoted correspondingly the critical distances $D_n$. Of course, the greater the value of *n*, the greater the biological damage. (Thomas et al, 2005) assumed that *n=2* would cause significant marine mortality, especially in phytoplankton, which lives near the surface, as it needs light to make photosynthesis. However, we remind the reader that we have used effective biological *fluences*, while they used *fluxes*.

Also, as our work is concerned with several eons, we point out that having a greater instantaneous damage caused by the GRB (as seen in the former subsection) by no means imply a greater final or definitive damage, because we should also consider the solar UV instantaneous damage which the species is used to live with. Obviously, these two magnitudes increased with the reduction of the oxygen content, as seen in Table 3, which also shows a set of critical distances.

| $O_2$ content (pal) | $E^*_{ad}$ | $E^*_{Sun}$ (W/m$^2$) | $D_2$ (kpc) | $D_4$ (kpc) | $D_8$ (kpc) | $D_{16}$ (kpc) |
|---|---|---|---|---|---|---|
| 1 | 0.005028 | 0.009158 | 4.45 | 3.14 | 2.22 | 1.57 |
| $10^{-1}$ | 0.01006 | 0.01410 | 5.07 | 3.58 | 2.53 | 1.79 |
| $10^{-2}$ | 0.042528 | 0.09561 | 4.00 | 2.83 | 2.00 | 1.42 |
| $10^{-3}$ | 0.084888 | 1.9610 | 1.25 | 0.88 | 0.63 | 0.44 |
| $10^{-4}$ | 0.1283 | 33.4535 | 0.37 | 0.26 | 0.18 | 0.13 |
| $10^{-5}$ | 0.1719 | 50 | 0.35 | 0.25 | 0.17 | 0.12 |

**Table 3: Critical distances for the UV-flash to double the effective biological fluence caused by the solar UV. Note:** 1kpc = 3.1 x 10$^{19}$ m

**IV - General discussion**

From above section we can infer:

1) Phanerozoic, Late- and Mid-Proterozoic biospheres would be, in the short-term, the more stressed by the UV flash of a GRB coming from several kpc. Note that even to potentially cause a damage 16 times greater than that of the solar UV (last column), the critical distances for those eras are namely the estimated for the "last typical nearest burst" potentially striking Earth in the last billion years, which is considered to have been emitted at one or two kpc from our planet. Thus, it could be interesting to include this effect in the very detailed modeling of long-term effects of a GRB on Phanerozoic eon (Thomas et al, 2005).
2) For a given distance to the GRB progenitor, the ecosystems living in ~10$^{-1}$ pal atmosphere would feel a greater immediate stress. This roughly corresponds to a Mid-Proterozoic atmosphere.
3) **The last three rows of Table 3 indicate that the less protected ecosystems are not necessarily the more affected. They would certainly receive a higher biological fluence from a GRB, but they were somehow used to receive higher effective biological fluences from the everyday Sun. Therefore, it is appealing to think that during the Archean and Early Proterozoic the biota more resistant to radiations was on the land. Therefore, a GRB would affect it less than later biota used to live in milder photobiological environments, such as those from the Mid-Proterozoic up to the present.**

**V - Conclusions**

In this work we have estimated the biological importance of the UV-flash which a GRB would deliver on our planet's surface, considering different model atmospheres, potentially representing the main stages of the geological and biological evolution of our planet.

At first glance, one might think that the UV-flash is just a transient short-term effect, because a typical long burst only lasts from 10 seconds to a minute, thus affecting directly only one hemisphere during a small time. On above grounds, this effect is not usually considered. However, although the irradiation time is short, mortality of surface organisms can occur later as a result of irreversible DNA damages and other harmful biochemical effects in exposed cells. Phytoplankton, obliged to live in the light-exposed zone of aquatic ecosystems, would certainly be a primary target. We have defined an effective biological fluence to have a better account of biological damage, and we have shown that the brief UV-flashes from GRB's emitted at distances of the order of kpc are capable of doubling the solar UV effective biological fluence of a *whole day* (Table 3). The consequent damage on phytoplankton and other surface groups could be transmitted trophically to other ecological niches, creating longer-term detrimental effects on the biota. Actually, at the moment there is the hypothesis that the minor marine extinction of tropical mollusks around two million years ago was caused by a similar mechanism: radiation coming from a nearby supernova firstly affected phytoplankton and later this was trophically transmitted to low-mobility groups such as bivalve mollusks (Benítez, Maiz-Apellaniz & Canelles, 2002). The aquatic ecosystems have always hosted an important fraction of global primary productivity in our planet. For instance, today the ocean accounts for about half of global primary productivity and, accordingly, half of $CO_2$ fixation and $O_2$ evolution (Behrenfeld et al, 2005). Thus, even modest effects on oceanic productivity could have major repercussions throughout the whole marine food chain and, through atmospheric effects, the global climate (Thomas, 2009).

We also would like to comment on another point now. Most times we expect Nature to respond linearly to small perturbations. However, ecologists are detecting abrupt changes in ecosystems due to small changes in a given parameter. This is quite analogous to phase transitions in physical systems. Seemingly, there are thresholds in many ecosystems, which determine the rapid shift to another state (van Nes & Scheffer, 2004). There is some empirical evidence showing that several aquatic ecosystems have undergone relatively rapid phase shifts during the past century (Scheffer et al, 2001). Therefore, under some circumstances, a short-term perturbation can provoke a rapid phase shift of an entire ecosystem. We are now checking whether a brief UV-flash can cause abrupt phase transitions in aquatic ecosystems, and in some cases the answer has been affirmative. We will present this in a forthcoming paper.

We are aware of the fact that potential long-term effects, such as alterations of the atmospheric chemistry with its subsequent effects, should have priority when modeling the influence of a GRB on our planet. However, as new developments make the modeling more accurate, the inclusion of the UV-flash as an initial scenario to infer the long-term effects could shed more light on what a GRB would cause in our planet. Perhaps interesting surprises are waiting for the model builders.

**Acknowledgements**

The authors would like to acknowledge the financial support from FAPESP (Brazil), in the form of fellowships, and CAPES (Brazil), which has funded a joint program USP (Brazil)-Universidad Central de Las Villas, Santa Clara (Cuba). This work has received a very important benefit from this support.